\documentclass[runningheads]{llncs}
\usepackage{graphicx}
\usepackage{booktabs}
\usepackage{amsmath}
\usepackage{amssymb}
\usepackage{url}
\usepackage[hidelinks]{hyperref}
\usepackage{xcolor}
\usepackage{multirow}
\usepackage{adjustbox}
\usepackage{tikz}
\usepackage{colortbl}
\usepackage{cite}
\usepackage{xcolor}
\usepackage{tcolorbox}  
\tcbuselibrary{breakable, skins}  
\usepackage{wrapfig}
\usetikzlibrary{decorations.pathreplacing}
\usepackage{array}    
\usepackage{tcolorbox}[most]
\newcolumntype{C}[1]{>{\centering\arraybackslash}p{#1}}
\usepackage{etoolbox}
\apptocmd{\thebibliography}{
  \scriptsize
  \setlength{\itemsep}{0pt}
  \setlength{\parskip}{0pt}
}{}{}
\makeatletter
\renewcommand\paragraph{\@startsection{paragraph}{4}{\z@}%
  {1ex \@plus 0.5ex \@minus .2ex}%
  {-1em}%
  {\normalfont\normalsize\bfseries}}
\makeatother
\begin{document}

\title{Zoom, Don't Wander: Why Regional Search Outperforms Pareto Reasoning and Global Optimization in Budget-Constrained SBSE}
\titlerunning{Zoom, Don't Wander}

\author{Kishan Kumar Ganguly \and
Tim Menzies}

\authorrunning{K. K. Ganguly and T. Menzies}

\institute{North Carolina State University, Raleigh NC, USA\\
\email{kgangul@ncsu.edu, timm@ieee.org}}

\maketitle

\begin{abstract}
Traditional Search-Based Software Engineering (SBSE) assumes global search and full Pareto exploration
are essential. We offer the following negative result based on a study of over 100 Software Engineering (SE) optimization
tasks: {\bf ``zooming'' into promising regions is far more
effective than Pareto and global exploration under constrained evaluation budgets}. Our minimal greedy zoom method, EZR, runs three orders of magnitude faster than Pareto and global Bayesian methods, achieving higher statistical ranks and winning or tying in 84-89\% of datasets on equal budget. Even at one-fifth the evaluation budget, EZR wins or ties in 79-81\% of datasets. Surprisingly, despite never explicitly seeking a frontier, EZR matches or outperforms Pareto methods on their own coverage metrics (IGD, HV) at equal budgets. The explanation for this widespread failure is structural: across the datasets studied, Pareto-optimal solutions form a tiny, tight island concentrated in a compact region of decision space. Methods that wander waste their budgets outside this island.
Beyond efficiency, zooming yields small,
interpretable models, thus addressing concerns about black-box AI. 
By replacing global wandering with
greedy zooming, we make SBSE much faster, more explicable, and hence accessible to a wider audience.
SBSE practitioners and researchers should zoom, not wander.

\end{abstract}

\section{Introduction}

Software engineering optimization is inherently multi-objective:
release planning trades customer value against delivery
cost~\cite{bagnall2001next}, test suites balance coverage against
execution time~\cite{fraser2011evosuite}, defect prediction
simultaneously minimizes false alarms and missed
faults~\cite{tantithamthavorn2016automated}, and process simulation
models jointly optimize risk, cost, and schedule
deviation~\cite{menzies2025moot}. The Search-Based Software Engineering (SBSE) community's dominant
response to these problems is Pareto-based evolutionary
search~\cite{harman2012search, sorokin2025can, chenli2023weights, Li22}: run a multi-objective optimizer (e.g., NSGA-II \cite{deb2002fast} or SPEA2 \cite{zitzler2001spea2}), obtain a
frontier of trade-off options, and let the practitioner choose.
For large and complex configuration spaces, surrogate-assisted global Bayesian optimization is a natural recourse from the broader optimization literature ~\cite{hutter2011smac, bischl2023hyperparameter} .

Both are expensive. Our landscape analysis shows that Pareto solutions are rare ($\approx$\textbf{0.6\%} of configurations), concentrated in decision space (\textbf{85\%} of datasets), and clustered near the low-distance-to-ideal region of objective space (\textbf{88\%} of datasets). This raises a natural question:
\begin{quote}
{\bf RQ0:} 
\textit{Is expensive frontier exploration and global coverage worth their evaluation cost under realistic SE budgets?}
\end{quote}

To answer \textbf{RQ0}, this paper compares three search strategies across 100+ tasks from MOOT~\cite{menzies2025moot}: (1) \textbf{EZR}~\cite{Amiraliminimaldata,Menzies2025DataLight}, a minimal contrastive method; (2) \textbf{SMAC}~\cite{hutter2011smac}, a global Bayesian optimizer; and (3) \textbf{NSGA-II/SPEA2}~\cite{deb2002fast,zitzler2001spea2}, Pareto-diversity algorithms. This leads
to the following negative 
result:

\begin{tcolorbox}[
    enhanced, colback=gray!5, colframe=gray!40,
    boxrule=0.4pt, arc=2pt,
    left=3pt, right=3pt, top=2pt, bottom=2pt]
  \textbf{Negative Result:} Under realistic SE budgets, Pareto exploration
  and global Bayesian search do not justify their costs.
  \end{tcolorbox}

While this is a {\bf negative results} paper, we assert  that it is very good news for users and researchers of SBSE
algorithms for two reasons.

Firstly, there is the issue of {\bf efficiency}.
As shown below, greedy search is orders of magnitude faster than algorithms seeking global coverage. Since  it performs as well as these expensive alternatives, our results mean that SBSE methods can be explored and deployed much faster to a much wider audience.

\begin{wrapfigure}[13]{r}{.5\textwidth}
\includegraphics[width=.5\textwidth]{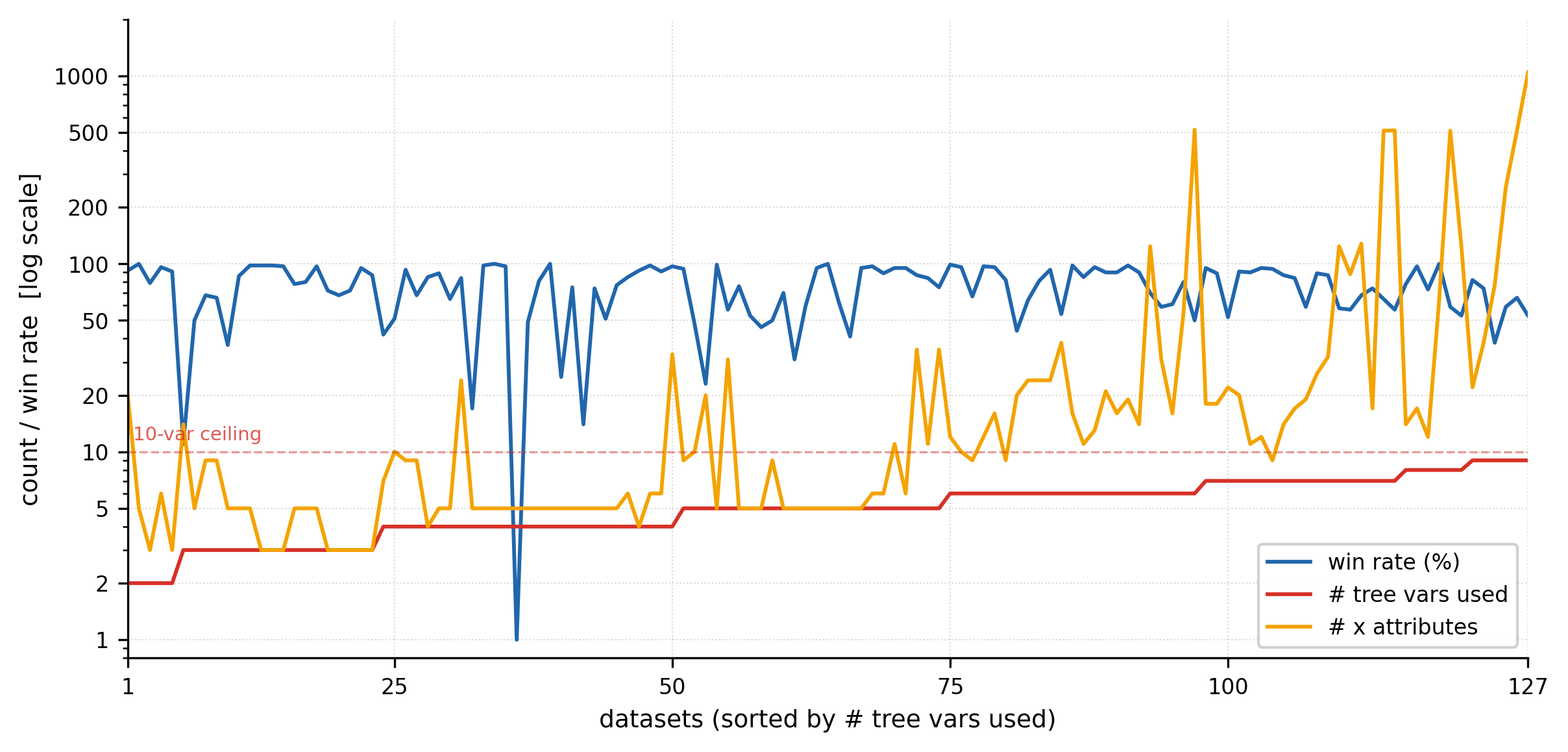}
\caption{Red= \#variables used in model;
yellow= \#variables in dataset features;
blue= model efficacy (defined in \S\ref{eval},   so {\em larger} numbers are {\em better}).
For an example of one of these models, see Figure~\ref{fig:ezr-tree-output}.~\\
}\label{fig:treecomplexity}
\end{wrapfigure}Secondly, \textbf{Interpretability} is another key issue. Our greedy search produces small and interpretable models. Figure~\ref{fig:treecomplexity}
shows the efficacy of models (blue) built from 100+ MOOT data sets, using a few variables (red) from data sets with 3-1000
variables (yellow). 
Note that efficacy often remains near the maximum (100) and almost unaffected by how few variables we use. 
Hence, at least for the data used in this study, tiny models are enough to explain complex optimization problems. This addresses a major concern raised by Rudin~\cite{rudin2019stop}, who argues that simple, interpretable models are required for high-stakes decisions.

 Overall, this study makes the following contributions:
 \begin{enumerate}

\item We show the relative efficacy of focused regional search with EZR versus Pareto methods and global Bayesian Optimization. On constrained budget, EZR wins/ties both in \textbf{84--89\%} of cases, obtaining best solutions in 75\% cases (vs 61-68\%) and beats Pareto methods on their own frontier metrics.
\item Even when given \textbf{5$\times$} more budget, Pareto and Global Bayesian search ties/loses to EZR (\textbf{79\%} of cases), which runs \textbf{2--3 orders of magnitude} faster.
\item Pareto solutions are rare ($\approx0.6\%$) and cluster in both decision (85\% of datasets) and objective space near the ideal values (low D2H, 88\% of datasets).
\item \textbf{Replication package:} \url{https://github.com/kkganguly/ParetoMyth}.
\end{enumerate}

\section{Background}

The  efficacy and interpretability benefits boasted above are  somewhat irrelevant
 if greedy search does not match the solution quality of established Pareto methods. This section argues that it can, by showing that (a)~When seeking one best solution, standard Pareto metrics measure the wrong thing for budget-constrained practitioners, and (b)~the structure of SE objective spaces favors zoom methods over frontier-wide exploration.

\vspace{-0.3cm}
\subsection{The Evaluation Bottleneck}
Both Pareto exploration and global search share the same fundamental strategy: wander the space broadly. The standard justification is that Pareto methods promise trade-off options spanning all preference weights and  global methods promise escape from local optima. Both assumptions embed the same implicit cost that deserves scrutiny: that evaluation budgets are realistically large enough to make broad exploration worthwhile.

In practice, the bottleneck is candidate evaluation cost, not the optimizer~\cite{nair2018flash,Menzies2025DataLight}. For example, compiler flag selection requires 35--135 minutes per evaluation~\cite{nair2018flash}, and Nair et al.\ explicitly note that ``exploring more than a handful of configurations is usually infeasible due to long benchmarking time''~\cite{nair2018flash}; similarly, expert elicitation yields roughly ten labels per session~\cite{valerdi2010heuristics,easterby1980design}. In such settings, overnight runs or quarterly efforts cap realistic budgets well below 1,000 evaluations (e.g. 100-200 evaluations). Yet, SBSE methods such as NSGA-II routinely assume populations of 100 across hundreds of generations~\cite{deb2002fast, chenli2023weights}. The right question is therefore not \emph{which optimizer wins given unlimited budget}, but \emph{which wins within what practitioners can actually afford}.

\vspace{-0.3cm}
\subsection{The Practitioner Selection Problem}
\begin{wrapfigure}{r}{0.4\textwidth}
\vspace{-12pt} 
\begin{tcolorbox}[
  enhanced,
  colback=gray!5,
  colframe=gray!40,
  boxrule=0.4pt,
  arc=2pt,
  left=3pt, right=3pt, top=2pt, bottom=2pt,
  title={\scriptsize\textbf{Pareto Metrics vs D2H}},
  fonttitle=\scriptsize,
  coltitle=black,
  attach boxed title to top left={yshift=-1.5mm, xshift=3mm},
  boxed title style={colback=gray!15, colframe=gray!40, boxrule=0.4pt}
]
\fontsize{7pt}{8pt}\selectfont
\setlength{\parskip}{2pt}

\textbf{GD}: convergence to true front; ignores spread.

\textbf{Spread}: diversity along front; ignores convergence.

\textbf{IGD}: captures both, but weights all preference regions
equally—including extreme trade-offs no practitioner selects.

\textbf{HV}: dominated objective-space volume; rewards
covering regions a practitioner would never choose.

\smallskip
\end{tcolorbox}
\vspace{-10pt}
\end{wrapfigure}

Pareto methods present practitioners with trade-off options; at the end, one solution is selected. The key question is therefore: under a constrained budget, which optimizer produces the best selectable outcome?

Without stated preferences, equal-weight aggregation is the theoretically grounded default, and the recommended first point to inspect is the solution closest to the ideal~\cite{hwang1981methods,zeleny1982multiple}. Standard indicators IGD and HV cannot measure the quality of a single best outcome, they reward coverage across all preference regions~\cite{li20}. We therefore use Distance-to-Heaven (D2H), the normalized Euclidean distance to the ideal, as our primary measure. Readers concerned this ignores coverage will find it addressed in RQ3. Under this criterion, exploration is costly. Every evaluation spent on Pareto diversity or global surrogate fitting is one fewer spent improving the best solution, which is a diversity tax that constrained budgets cannot absorb. Since the practitioner selects one solution and the budget is constrained, broad coverage is wasteful by construction.

It may be argued that reasonably increasing the budget may narrow the gap, but as we show next, it does not change the underlying structure of the problem, which is that Pareto-optimal solutions are concentrated in a small compact region close to the ideal solution. This suggests that a method zooming there should remain competitive regardless of budget, while global methods continue paying an exploration tax on every additional evaluation.

\vspace{-0.5cm} 
\subsection{Zoom Search Trajectories Cover the Pareto Front in SE Data}

\label{sec:landscape}
A natural objection to zoom methods is that they return a single good solution rather than a set of trade-off options. We address this by characterizing where the true Pareto front sits across our SE datasets. In multi-objective optimization this structure is not guaranteed: Pareto fronts can span the entire objective space, with solutions spread across extreme regions~\cite{hwang1981methods,zeleny1982multiple}. We extract the true Pareto front from each of our 100+ SE datasets via exhaustive non-domination and compute D2H for every configuration. To test whether Pareto solutions occupy a distinct low-D2H region, we compare their D2H distribution against non-Pareto solutions using a Mann-Whitney U test, which is a non-parametric rank test that requires no distributional assumptions and report effect size as rank-biserial correlation (rb), where rb $<$0 means Pareto solutions rank lower in D2H. We apply the same test to pairwise decision-space distances to check whether Pareto solutions are also spatially clustered:
\vspace{-0.4cm}
\begin{center}
\colorbox{gray!10}{%
\resizebox{\linewidth}{!}{%
\begin{tabular}{lr}
Median Pareto fraction of configurations & 0.6\% configurations \\
Pareto D2H $<$ non-Pareto D2H (rb $<$ 0) & 94\% datasets \\
Pareto D2H $<$ non-Pareto D2H (significant, $p{<}0.05$, rb $<$ 0, median rb $= -0.89$) & 88\% datasets \\
Pareto tighter in decision space (significant, $p{<}0.05$, rb $<$ 0) & 85\% datasets \\
\end{tabular}}}
\end{center}
This is consistent with a structure not previously documented for SE configuration problems: across all 100+ datasets, Pareto solutions are not only rare but doubly concentrated. They occupy a tight cluster in decision space while simultaneously sitting in the near-ideal region of objective space, rather than being dispersed across it as multi-objective theory permits~\cite{hwang1981methods,zeleny1982multiple}.

\paragraph{Implication.} A zoom method searching for low-D2H solutions will naturally accumulate near-Pareto solutions along its trajectory. Non-domination filtering over that trajectory should therefore yield practical trade-off coverage without any explicit diversity mechanism. \textbf{RQ3} tests this prediction: does EZR’s search trajectory provide reasonably good trade-off coverage
compared to methods explicitly designed for frontier exploration?



\section{Methodology}

\noindent The previous section justified our central question: \textbf{at realistic constrained budgets (50-200 evaluations in our study), does pareto and global exploration provide sufficient benefit to justify its use?}
\subsection{Research Questions}
\begin{itemize}
\item \textbf{RQ1 (Solution Quality):} Under constrained evaluation budgets, do Pareto methods or global Bayesian search outperform a minimal zoom method in solution quality, at equal or 5 times higher budget, enough to justify their runtime cost?
\item \textbf{RQ2 (Budget Sensitivity):} Do Pareto methods and global Bayesian search converge to better solutions as budget increases, or do they plateau early while EZR maintains its advantage?
\item \textbf{RQ3 (Trade-off Coverage):} Can a zoom method's search trajectory provide useful trade-off options, and how does the resulting frontier compare to that of methods explicitly designed for Pareto exploration?
\end{itemize}

\subsection{Algorithms}
\begin{wrapfigure}{r}{0.5\textwidth}
\vspace{-9pt}
\centering
\scriptsize
\textbf{Algorithm Taxonomy}\\[5pt]
\begin{tabular}{@{}p{1.4cm}p{1.2cm}p{3.4cm}@{}}
\toprule
\textbf{Category} & \textbf{Algo.} & \textbf{Description} \\
\midrule
\textit{Greedy}  
  & EZR      & Greedy exploitation of local promising regions \\[4pt]
\textit{Global}  
  & SMAC     & Bayesian search via Random Forest surrogate \\[4pt]
\multirow{2}{*}{\textit{Pareto}}  
  & NSGA-II  & Non-dominated sorting with crowding distance \\[2pt]
  & SPEA2    & Archive-based strength Pareto evolution \\[4pt]
\textit{Baseline}  
  & Random   & Uniform random exploration; zero learning \\
\bottomrule
\end{tabular}
\begin{tikzpicture}[scale=0.8, every node/.style={font=\scriptsize}]
  \draw[black, thick] (1.4,1.60) -- (5.8,1.60);
  \foreach \x in {1.4, 2.2, 3.4, 5.8}
    \filldraw[black] (\x, 1.60) circle (2.5pt);
  \node[font=\scriptsize, anchor=east] at (1.2, 1.60) {EZR};
  \draw[black!70, thick, dashed] (3.4,1.20) -- (5.8,1.20);
  \foreach \x in {3.4, 5.8}
    \filldraw[black!70] (\x, 1.20) circle (2.5pt);
  \node[font=\scriptsize, anchor=east] at (1.2, 1.20) {SMAC};
  \draw[black!50, thick] (3.4,0.80) -- (5.8,0.80);
  \foreach \x in {3.4, 5.8}
    \filldraw[black!50] (\x, 0.80) circle (2.5pt);
  \node[font=\scriptsize, anchor=east] at (1.2, 0.80) {NSGA-II};
  \draw[black!50, thick, dashed] (3.4,0.45) -- (5.8,0.45);
  \foreach \x in {3.4, 5.8}
    \filldraw[black!50] (\x, 0.45) circle (2.5pt);
  \node[font=\scriptsize, anchor=east] at (1.2, 0.45) {SPEA2};
  \draw[->, thick, gray!50] (1.2,-0.55) -- (6.5,-0.55);
  \foreach \x/\lbl in {1.4/50, 2.2/100, 3.4/200, 5.8/1000} {
    \draw[gray!60] (\x, -0.48) -- (\x, -0.62);
    \node[below=3pt, gray!70] at (\x, -0.52) {\lbl};
  }
  \draw[black!30, thick] (3.4,-0.02) -- (5.8,-0.02);
  \foreach \x in {3.4, 5.8}
    \filldraw[black!30] (\x, -0.02) circle (2.5pt);
  \node[font=\scriptsize, anchor=east] at (1.2, -0.02) {Random};
  
  \draw[gray!40, thin] (6.1, 1.45) -- (6.1, 1.75);
  \node[font=\scriptsize\itshape, anchor=west] at (6.15, 1.60) {Greedy};
  \draw[gray!40, thin] (6.1, 1.05) -- (6.1, 1.35);
  \node[font=\scriptsize\itshape, anchor=west] at (6.15, 1.20) {Global};
  \draw[gray!40, thin] (6.1, 0.30) -- (6.1, 0.95);
  \node[font=\scriptsize\itshape, anchor=west] at (6.15, 0.62) {Pareto};
  \draw[gray!40, thin] (6.1, -0.33) -- (6.1, 0.09);
  \node[font=\scriptsize\itshape, anchor=west] at (6.15, -0.02) {Baseline};
\end{tikzpicture}

\caption{Algorithm Budgets and Taxonomy}
\label{fig:algo-taxonomy}
\vspace{-0.15cm}
\end{wrapfigure}
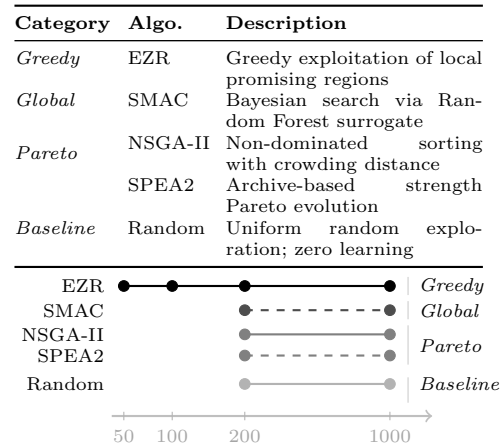~\vspace{-0.5cm}\subsubsection{Pareto Frontier Methods:}

NSGA-II~\cite{deb2002fast} is the one of the most widely cited multi-objective
evolutionary algorithm in SBSE~\cite{harman2012search}. As per Colanzi et al., it is the second most widely used optimizers in SBSE over the past decade \cite{colanzi2020symposium}. SPEA2~\cite{zitzler2001spea2}
provides an archive-based alternative with strength fitness and
$k$-NN density estimation. Including both ensures results are not
artifacts of NSGA-II's specific crowding-distance mechanism.
Both are run at 200 (equal-budget comparison) and
1000~evaluations ($5\times$ advantage).

\subsubsection{Global Bayesian Search:}

SMAC~\cite{hutter2011smac} fits a Random Forest surrogate over
the decision space and uses Expected Improvement to select
evaluations, optimising D2H directly. It is included because it
represents the mainstream alternative to Pareto methods for
practitioners seeking a single good balanced solution rather than
a frontier. A prior study on the same MOOT benchmark~\cite{Menzies2025DataLight} found it competitive with or better than TPE~\cite{bergstra11TPE} and DEHB~\cite{awad2021dehb}, making it a reasonable choice among global Bayesian methods for this comparison. 
\vskip -0.3cm
\subsubsection{Random Sampling Baseline:}
\vskip -0.3cm

Random sampling serves as an essential baseline representing zero algorithmic sophistication and pure uniform exploration without learning or adaptation. Arcuri and Briand~\cite{arcuri2011practical} demonstrated that random search often provides surprisingly competitive performance in SBSE, making it an essential baseline for any optimization comparison. We include Random at two budgets (200 and 1000 evaluations) to baseline with other optimizers' at similar budgets.
\vskip -0.3cm

\vspace{-0.2cm}
\subsubsection{Regional Search:}
\label{sec:ezr}
We use a minimal contrastive regional search, EZR \cite{Menzies2025DataLight, Amiraliminimaldata}. It maintains a \textbf{Best} set ($\sqrt{N}$ solutions with
lowest D2H found so far) and a \textbf{Rest} set (near-good
solutions displaced from Best as better ones are discovered).
At each step, EZR randomly samples 128~candidates from the
unlabeled pool and returns the first candidate closer to Best's
decision-space centroid than to Rest's centroid:
\begin{equation}
x_{\text{next}} = \text{first } x \in \mathcal{U}[128]
  \text{ s.t. } d_X(x,\,\mu_\mathcal{B}) < d_X(x,\,\mu_\mathcal{R})
\label{eq:ezr}
\end{equation}
where $d_X$ is normalized Euclidean distance over X-columns,
and $\mu_\mathcal{B}$, $\mu_\mathcal{R}$ are the centroids of
Best and Rest respectively. EZR's sampling of 128 candidates occurs entirely in the unlabelled decision space. Evaluating the $x$-distance of 128 candidates is computationally trivial and consumes zero objective-function evaluation budget. Only the single selected candidate $x_{\text{next}}$ is actually evaluated to find its $y$-consequences, meaning EZR strictly consumes exactly one budget unit per iteration. 

Crucially, Best is bounded at $\sqrt{N}$ solutions; when a
new solution enters, the worst incumbent is demoted to Rest.
The combined Best+Rest set therefore captures the full search
trajectory: Best holds the current best solutions and Rest accumulates
near-good solutions from earlier iterations when the search was
exploring slightly different parts of the good region. At
termination, a frontier can be constructed by non-domination over Best+Rest, which we utilize in RQ3.

We evaluate EZR at four budgets (50, 100, 200, and 1000 evaluations) to assess how quickly regional search converges relative to wandering methods, and to establish the minimum budget at which EZR's advantage over Pareto and global methods becomes apparent.

\textbf{Why EZR:} Among zooming methods (LITE~\cite{Menzies2025DataLight},
SWAY~\cite{chen2018sampling}, etc.), EZR is an ideal representative
because its simplicity isolates zooming as the sole driver.
It has no surrogate, no Pareto reasoning, no complex structures, and its
success despite minimal sophistication strengthens our negative result.
\vspace{-0.2cm}

\section{Experimental Design}
\subsection{Datasets: The MOOT Benchmark Repository}
We evaluate on MOOT (Multi-Objective Optimization Testing)~\cite{menzies2025moot}, a repository consolidating over 100 SE optimization tasks drawn from dozens of peer-reviewed publications across multiple years of community research. Rather than selecting benchmarks to favor any particular method, MOOT aggregates what the SE community has actually studied. Hence, its diversity is inherited from the literature, not engineered for this evaluation.

Tasks span multiple SE domains such as: \textbf{software configuration} (e.g., system performance tuning across 25 software systems and 12 PromiseTune tasks \cite{chen2026promisetune}); \textbf{project health prediction} (GitHub health forecasting across closed issues, PRs, and commits); \textbf{software process models} (effort and risk estimation including POM3 and XOMO variants); \textbf{feature models} (Scrum and FM constraint satisfaction problems), software testing tasks and \textbf{cross-domain analytics} (spanning behavioral, financial, and health). Each dataset provides input configurations x and objective values y, with +/- notation indicating maximize/minimize directives. Full dataset details appear in Table~\ref{combinedtable}.

\begin{table*}[!t]
\caption{Summary of MOOT datasets. ``Features'' = number of input variables.}
\vspace{-0.2cm}
\label{combinedtable}
\begingroup
\renewcommand{\arraystretch}{0.5}
\setlength{\tabcolsep}{2pt}
\begin{adjustbox}{max width=\textwidth}
\scriptsize
\begin{tabular}{@{}rp{1.5cm}>{\raggedright\arraybackslash}p{3.8cm}
p{2.8cm}p{1.2cm}p{1.2cm}p{2.4cm}@{}}
\toprule
\# & Type & Datasets & Objective & Features & Rows & Cited \\
\midrule
37 & Config Tuning & SS-A to SS-X, billing10k, 7z, BDBC, HSQLDB, LLVM, PostgreSQL, dconvert, deeparch, exastencils, javagc, redis, storm, x264 & System \& Performance optimisation & 3--88 & 197--167k & \cite{Amiraliminimaldata,menzies2025the,lustossa2024isneak,senthilkumar2024can,lusstosa2025less,nairMSR18,nair2017using, chen2026promisetune,chen2025accuracy} \\
\midrule
1 & Cloud & HSMGP\_num & Hazardous SW & 14 & 3,457 & \cite{Amiraliminimaldata,menzies2025the,chen2025accuracy,senthilkumar2024can} \\
1 & Cloud & Apache AllMeasurements & Server performance & 9 & 192 & \cite{Amiraliminimaldata,menzies2025the,chen2025accuracy,senthilkumar2024can} \\
1 & Cloud & SQL AllMeasurements & DB tuning & 39 & 4,654 & \cite{Amiraliminimaldata,menzies2025the,senthilkumar2024can} \\
1 & Cloud & X264 AllMeasurements & Video encoding & 16 & 1,153 & \cite{Amiraliminimaldata,menzies2025the,senthilkumar2024can} \\
7 & Cloud & (rs--sol--wc)* & Misc config & 3--6 & 196--3,840 & \cite{Amiraliminimaldata,menzies2025the,senthilkumar2024can,lusstosa2025less,nairMSR18} \\
\midrule
35 & Project Health & Health-ClosedIssues,-PRs,-Commits & Health prediction & 5 & 10,001 & \cite{Amiraliminimaldata,menzies2025the,senthilkumar2024can,lusstosa2025less,lustosa2024learning} \\
\midrule
3 & Scrum & Scrum1k,10k,100k & Feature model config & 124 & 1k--100k & \cite{Amiraliminimaldata,menzies2025the,lusstosa2025less,lustossa2024isneak} \\
8 & Feature Models & FFM-*,FM-* & Variables/constraints & 128--1044 & 10,001 & \cite{Amiraliminimaldata,menzies2025the,lusstosa2025less,lustossa2024isneak} \\
\midrule
1 & SW Process & nasa93dem & Effort/defects/LOC & 24 & 93 & \cite{menzies2025the,senthilkumar2024can,lusstosa2025less,lustosa2024learning} \\
1 & SW Process & coc1000 & Risk/effort/experience & 20 & 1,001 & \cite{Amiraliminimaldata,menzies2025the,senthilkumar2024can,lustosa2024learning,chen2018beyond} \\
1 & SW Process & pom3d & Idle/completion/cost & 9 & 501--20k & \cite{menzies2025the,senthilkumar2024can,lusstosa2025less,lustosa2024learning,lustossa2024isneak} \\
\midrule
3 & Behavioral & student\_dropout, HR-attrition, player\_stats & Behavioral patterns & 26--55 & 82--17k & \cite{abdullah0a_student_dropout_analysis_prediction_2025,die9origephit_fifa_wc_2022_complete_2025,pavansubhasht_ibm_hr_analytics_attrition_2025} \\
2 & Financial & home\_data, Telco-Churn & Financial prediction & 19--77 & 1,460--20k & \cite{blastchar_telco_customer_churn_2025,dansbecker_home_data_for_ml_course_2025} \\
2 & Health & Life\_Expectancy, hospital\_Readmissions & Health prediction & 20--64 & 2,938--25k & \cite{dansbecker_hospital_readmissions_2025,kumarajarshi_life_expectancy_who_2025} \\
2 & RL & A2C\_Acrobot, A2C\_CartPole & RL tasks & 9--11 & 224--318 & \\
5 & Sales & accessories, dress-up, Marketing, socks, wallpaper & Sales prediction & 14--31 & 247--2,206 & \cite{jessicali9530_animal_crossing_new_horizons_nookplaza_dataset_2021,jackdaoud_marketing_data_2022,syedfaizanalii_car_price_dataset_cleaned_2025} \\
3 & Misc & auto93, Car\_price, Wine\_quality & Miscellaneous & 5--38 & 205--1,600 & \cite{Amiraliminimaldata,menzies2025the,senthilkumar2024can,lusstosa2025less,lustosa2024learning} \\
\midrule
\textbf{114} & \textbf{Total} & & & & & \\
\bottomrule
\end{tabular}
\end{adjustbox}
\endgroup
\end{table*}

\subsection{Evaluation Metrics}\label{eval}

\textbf{D2H} (primary, RQ1--RQ2) is the normalized Euclidean distance
to the ideal point where all objectives are simultaneously optimal:
\begin{equation}
\text{D2H}(\mathbf{y}) =
\sqrt{
\frac{1}{M}
\sum_{i=1}^{M}
\left(\hat{y}_i - h_i\right)^2
}
\label{eq:d2h}
\end{equation}
Lower is better; the mean formulation ensures comparability across
datasets with different numbers of objectives.

\textbf{Validation Regret} (RQ2) measures the relative gap to the
best achievable solution at evaluation $t$:
\begin{equation}
\text{regret}(t) = \frac{\text{D2H}^{\text{best}}(t) -
\text{D2H}^{*}}{\text{D2H}^{*}}
\label{eq:regret}
\end{equation}
where $\text{D2H}^{*}$ is the minimum D2H over the exhaustively
labeled dataset, making regret comparable across datasets with
different absolute D2H scales.

\textbf{Frontier coverage metrics} (RQ3): True IGD, True GD, HV,
and Spacing, computed against the exhaustive true Pareto front.
These metrics are appropriate for RQ3 because it asks a different
question from RQ1-RQ2: not which method finds the best balanced
solution, but whether EZR's frontier covers all trade-off positions
that a practitioner with any preference weight could want. True IGD
and HV penalize any method that misses any region of the true front
regardless of preference weight, providing the strongest possible
test of the diversity objection. Critically, all metrics are
computed against the true front extracted by exhaustive
non-domination over the fully labeled dataset, so scores reflect
coverage of wherever the SE front actually lies rather than
penalizing methods for not covering synthetic uniform distributions.

We perform statistical comparisons via \textbf{Scott-Knott} recursive bi-clustering method~\cite{Scott1974}. It clusters optimizers
per dataset by recursively splitting ranked distributions only where
bootstrap significance and Cliff's Delta effect size
both confirm a meaningful difference, accounting for both central
tendency and variance. Tier~1 denotes the best group, Tier-2 second best etc. In our experiments, each
algorithm runs 20 independent trials.

\section{Results}

\subsection{RQ1: Solution Quality}
\label{sec:rq1}

Table~\ref{tab:skott} reports Scott-Knott tier counts and
Win/Tie/Loss comparisons. Win/\-Tie/\-Loss counts follow the pairwise practice recommended by Demšar \cite{demvsar2006statistical}: the cheaper method is the reference, so a Win means EZR-200 reached a better Scott-Knott tier on that dataset, a Tie means both share the same tier, and a Loss means the opponent reached a higher tier. This perspective is intentional: we ask whether the cheaper method holds its own, not the reverse.

\begin{table}[t!]
\centering
\caption{Scott-Knott summary across 114 SE datasets.
\colorbox{green!20}{Green} = best per row per zone. Win/Tie/Loss:
reference is always the cheaper method; W+T $\geq$ 50\% means it
holds its own or better.}
\vspace{-0.2cm}
\label{tab:skott}
\setlength{\tabcolsep}{1.4pt}
\scriptsize
\resizebox{\textwidth}{!}{%
\begin{tabular}{lrrr|rrrr||rrrr|r}
\toprule
& \multicolumn{7}{c||}{\textbf{Practical Zone (200 evals)}}
& \multicolumn{5}{c}{\textbf{5$\times$ Budget Zone (1000 evals)}} \\
\cmidrule(lr){2-8}\cmidrule(lr){9-13}
& \textbf{EZR} & \textbf{EZR} & \textbf{EZR}
& \textbf{SMAC} & \textbf{NSGA-II} & \textbf{SPEA2} & \textbf{Rand}
& \textbf{EZR}
& \textbf{SMAC} & \textbf{NSGA-II} & \textbf{SPEA2} & \textbf{Rand} \\
& \textbf{50} & \textbf{100} & \textbf{200}
& \textbf{200} & \textbf{200} & \textbf{200} & \textbf{200}
& \textbf{1k}
& \textbf{1k} & \textbf{1k} & \textbf{1k} & \textbf{1k} \\
\midrule
Tier 1 (\%)
  & 26 & 50 & \cellcolor{green!20}\textbf{75}
  & 68 & 61 & 63 & 46
  & \cellcolor{green!20}\textbf{98}
  & 86 & 92 & 88 & 81 \\
Tier 1--2 (\%)
  & 69 & 89 & \cellcolor{green!20}\textbf{98}
  & 94 & 89 & 89 & 82
  & \cellcolor{green!20}\textbf{100}
  & 98 & 99 & 96 & 97 \\
Tier 3+ (\%)
  & 31 & 11 & \cellcolor{green!20}\textbf{2}
  &  6 & 11 & 11 & 18
  & \cellcolor{green!20}\textbf{0}
  &  2 &  1 &  4 &  3 \\
Runtime (s)
  & \cellcolor{green!20}\textbf{1.1} & \cellcolor{green!20}\textbf{2.1} & \cellcolor{green!20}\textbf{5.2}
  & 5060 & 1396 & 1420 & 0.3
  & \cellcolor{green!20}\textbf{0.25}
  & 36229& 7800 & 8100 & 2.1 \\
\midrule
Evals
  & 50 & 100 & 200
  & 200 & 200 & 200 & 200
  & 1k & 1k & 1k & 1k & 1k \\
\bottomrule
\end{tabular}}

\vspace{2pt}
\resizebox{\textwidth}{!}{%
\begin{tabular}{llrrrr||llrrrr}
\toprule
\multicolumn{6}{c||}{\textbf{Practical Zone (200 evals)}} &
\multicolumn{6}{c}{\textbf{5$\times$ Budget Zone}} \\
\textbf{Ref} & \textbf{Opponent} & \textbf{W} & \textbf{T} & \textbf{L} & \textbf{W+T vs L} &
\textbf{Ref} & \textbf{Opponent} & \textbf{W} & \textbf{T} & \textbf{L} & \textbf{W+T vs L} \\
\midrule
EZR-200  & NSGA-II-200  & 36 & 61 & 17 & 97 vs 17 (\textbf{85\%}) &
EZR-200   & NSGA-II-1000 &  6 & 78 & 22 & 84 vs 22 (\textbf{79\%}) \\
EZR-200  & SPEA2-200    & 33 & 63 & 18 & 96 vs 18 (\textbf{84\%}) &
EZR-200   & SPEA2-1000   &  9 & 76 & 22 & 85 vs 22 (\textbf{79\%}) \\
EZR-200  & SMAC-200     & 22 & 79 & 13 & 101 vs 13 (\textbf{89\%}) &
EZR-200   & SMAC-1000    &  9 & 81 & 21 & 90 vs 21 (\textbf{81\%}) \\
EZR-200  & Rand-200     & 44 & 67 &  3 & 111 vs 3 (\textbf{97\%}) &
EZR-200   & Rand-1000    & 16 & 75 & 23 & 91 vs 23 (\textbf{80\%}) \\
Rand-200 & NSGA-II-200  & 14 & 67 & 33 & 81 vs 33 (\textbf{71\%}) &
Rand-1000 & NSGA-II-1000 &  5 & 86 & 15 & 91 vs 15 (\textbf{86\%}) \\
Rand-200 & SPEA2-200    & 14 & 63 & 37 & 77 vs 37 (\textbf{68\%}) &
Rand-1000 & SPEA2-1000   &  7 & 88 & 12 & 95 vs 12 (\textbf{89\%}) \\
Rand-200 & SMAC-200     &  4 & 74 & 36 & 78 vs 36 (\textbf{68\%}) &
Rand-1000 & SMAC-1000    & 11 & 84 & 16 & 95 vs 16 (\textbf{86\%}) \\
\bottomrule
\end{tabular}}
\vspace{-1pt}
\end{table}

\paragraph{Equal constrained budget: }
At 200 evaluations, EZR wins or ties NSGA-II in 85\% of datasets
and SPEA2 in 84\%, reaching Tier~1 in 75\% of datasets versus
61\% and 63\% respectively, while being about two orders of
magnitude faster. EZR also wins or ties SMAC in $89\%$ of the datasets while being three orders of magnitude faster. The random
baseline makes the structural failure of diversity maintenance
explicit: uniform sampling with zero algorithmic machinery wins
or ties NSGA-II-200 in 71\% of datasets and SPEA2-200 in 68\%,
and holds its own against SMAC-200 in 68\%. A method that does
nothing but sample randomly outperforms methods that deliberately
maintain diversity and fit surrogates, at the same budget. At constrained budgets, the
evaluations consumed by these mechanisms actively seek global coverage, hence displacing evaluations that would otherwise find
better solutions, consistent with the landscape structure
established earlier. EZR wins or ties
Random-200 in 97\% of datasets, confirming that directed
contrastive search adds value beyond mere sampling.

\paragraph{5$\times$ budget disadvantage:}
EZR at 200 evaluations wins or ties 79\% against NSGA-II-1000,
79\% against SPEA2-1000, and 81\% against SMAC-1000, at three
to four orders of magnitude lower runtime than its opponents.
The key diagnostic is the EZR-200 versus Random-1000 comparison:
EZR at 200 evaluations holds its own against Random at 1000
evaluations in 80\% of datasets. Since Random-1000 in turn matches
NSGA-II-1000 in 86\% of datasets and SPEA2-1000 in 89\%, the
improvement Pareto methods show at high budget is fully explained
by additional evaluations reaching the good region through
coverage, not by the diversity mechanism becoming useful. Any
method given enough evaluations will encounter the good region.
Despite budget disadvantage, EZR arrives there in many datasets at one-fifth the cost and orders of magnitude
faster.
\vskip -0.5cm
\paragraph{EZR at 1000 evaluations:}
EZR-1000 reaches Tier~1 in 98\% of datasets, the strongest result
in the table, confirming that EZR's contrastive search is not
inherently limited by design. The practical recommendation remains
EZR-200: the gain from 200 to 1000 evaluations is real but small
relative to the 5$\times$ cost increase.
\vspace{-0.1cm}
\begin{tcolorbox}[colback=gray!5, colframe=gray!40, boxrule=0.4pt,
  arc=1.5pt, left=4pt, right=4pt, top=2pt, bottom=1pt]
\small\textbf{RQ1.}
At equal budget, EZR wins or ties Pareto methods in 84–85\% of datasets and SMAC in 89\%, at orders of magnitude lower runtime. Random sampling alone matches these in 68–71\%, confirming that diversity maintenance and surrogate fitting displace useful evaluations. At 5× disadvantage, EZR-200 holds its own in 79–81\% against all opponents, proving high-budget gains reflect coverage not search mechanism, neither advantage justifies the runtime cost.
\end{tcolorbox}

\subsection{RQ2: Budget Sensitivity Analysis}
\label{sec:rq2}
\begin{wrapfigure}{r}{0.5\textwidth}
\vspace{-3pt}
\centering
\includegraphics[width=0.5\textwidth]{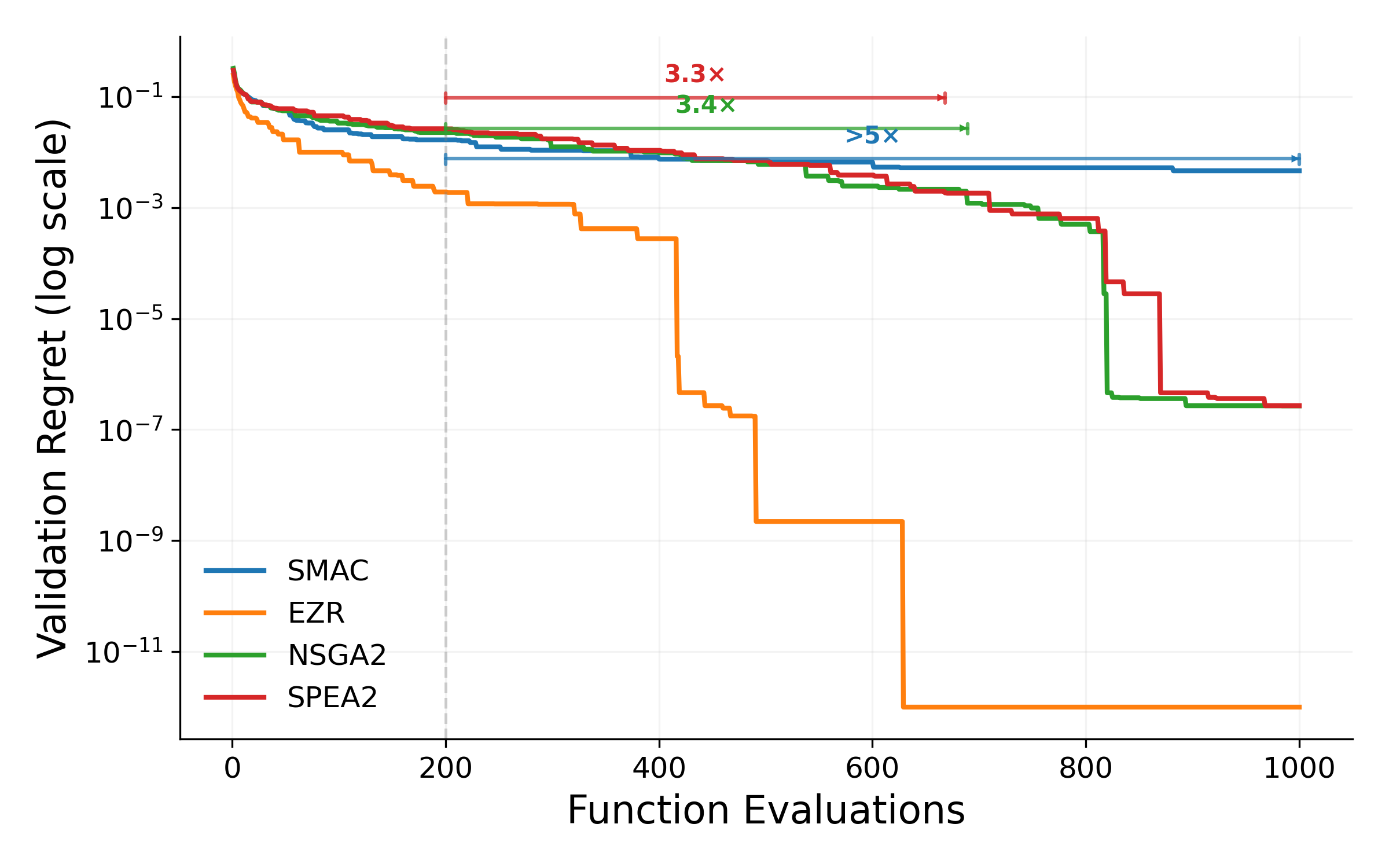}
\caption{Validation regret (log scale). Pareto methods need 3.3--3.4$\times$ budget vs EZR-200 to match
it; SMAC needs $>5\times$.}
\label{fig:budget_sensitivity}
\end{wrapfigure}

RQ1 established that EZR matches or beats Pareto methods and SMAC
at equal budget, and holds its own even at a 5$\times$ disadvantage.
RQ2 asks why: does EZR converge faster, do wandering methods improve
steadily with budget, and does additional budget eventually close
the gap? Figure~\ref{fig:budget_sensitivity} shows validation regret (Eq.~\ref{eq:regret}), which is the normalized gap to the best achievable D2H, versus evaluation budget for EZR, NSGA-II, SPEA2, and SMAC..

EZR drops steeply in the first 50-100 evaluations and reaches
a stable operating level by evaluation 150-200, confirming that
its practical-zone quality is achieved early. Beyond 200
evaluations EZR continues to improve, with a pronounced secondary
drop between evaluations 400-600 of several orders of magnitude
in median regret. This reflects EZR's search progressively
locating near-optimal solutions on more
datasets as it zooms into the good region. 


NSGA-II and SPEA2 improve through the practical zone but require
approximately 3.3-3.4$\times$ more evaluations to reach the regret level EZR achieves at 200
evaluations. Beyond that crossover both curves continue to
decline slowly and exhibit a sharp secondary drop at about 800 evaluation which flattens almost immediately. In spite of this drop, both remain several orders of magnitude above EZR's
regret by evaluation 1000.

SMAC is the slowest to converge. Its Random Forest surrogate
requires broad decision-space coverage before Expected Improvement
becomes informative. This makes early evaluations exploratory by
design. As shown in Figure~\ref{fig:budget_sensitivity}, SMAC does not reach
EZR's 200-evaluation regret until beyond 5$\times$ the budget.
This cold-start cost is structural: global Bayesian search cannot
exploit the concentrated good region in SE landscapes until it
has explored enough to trust its surrogate, by which
point EZR has converged.

\begin{tcolorbox}[colback=gray!5, colframe=gray!40, boxrule=0.4pt,
  arc=1.5pt, left=4pt, right=4pt, top=2pt, bottom=2pt]
\small\textbf{RQ2.} EZR reaches a stable operating level by 150--200 evaluations.
Pareto methods need 3.3--3.4$\times$ more to match it and SMAC
needs 5$\times$ due to cold-start overhead. No wandering method
approaches EZR's regret trajectory within practical SE budget.
\end{tcolorbox}

\subsection{RQ3: Frontier Quality and the Trade-off Options Objection}
\label{sec:rq3}
\vskip -0.2cm
RQ1 shows that diversity maintenance actively hurts at equal
budget and RQ2 confirms that additional budget buys coverage rather
than better search. The remaining objection is: \emph{Pareto methods
offer diverse trade-off options spanning all preferences, which D2H cannot
capture.} RQ3 addresses this using the standard Pareto
coverage metrics. 

EZR is a single-solution method, but at termination we apply
non-domination filtering over the combined Best+Rest trajectory to extract a frontier at no additional
evaluation cost. Because Best holds the current low-D2H region and
Rest captures near-good solutions from earlier iterations, the
resulting non-dominated set spans low D2H region of the true
front naturally, reaching near the Pareto front as per landscape argument. Table~\ref{tab:frontier} summarizes frontier coverage.

\begin{table}[t]
\caption{Frontier coverage across 114 datasets. True IGD and True GD:
  lower is better. HV: higher is better. SK\%T1 = fraction of datasets
  in the best Scott-Knott tier. \textbf{Bold} = best within each budget
  block.}
\label{tab:frontier}
\centering
\scriptsize
\setlength{\tabcolsep}{4pt}
\begin{tabular}{lrrrrrr}
\toprule
Method & T.IGD & T.GD & HV & Fr.\ sz & SK\%T1 (IGD) & SK\%T1 (HV) \\
\midrule
\multicolumn{7}{l}{\textit{Equal budget (200 evals)}} \\[1pt]
\textbf{EZR-200}      & \textbf{0.04} & \textbf{0.04} & \textbf{1.03} & \textbf{7.0} & \textbf{65\%} & \textbf{60\%} \\
NSGA-II-200           & 0.09          & 0.05          & 0.99          & 6.0          & 20\%          & 26\% \\
SPEA2-200             & 0.07          & 0.04          & 0.99          & 6.0          & 26\%          & 29\% \\
\midrule
\multicolumn{7}{l}{\textit{5$\times$ budget (1000 evals)}} \\[1pt]
\textbf{EZR-1000}              & \textbf{0.03}          & \textbf{0.01}          & \textbf{1.06}          & 8.1          & 90\%          & 87\% \\
\textbf{NSGA-II-1000} & \textbf{0.03} & \textbf{0.01} & \textbf{1.06} & 8.0 & \textbf{91\%} & \textbf{88\%} \\
SPEA2-1000            & 0.04          & 0.02          & 1.05          & 7.8          & 88\%          & 83\% \\
\bottomrule
\end{tabular}
\end{table}

\paragraph{Equal budget: }EZR achieves true IGD 0.04 versus 0.09 for NSGA-II-200 and 0.066
for SPEA2-200, and reaches Tier~1 on IGD in 65\% of datasets versus
20\% and 26\%. This is the starkest form of the negative result:
at equal budget, diversity maintenance does not build frontier
coverage, it prevents it. Every evaluation spent on diversity
maintenance is one fewer evaluation finding solutions near the
true front, as predicted by the landscape structure.
\paragraph{At 5$\times$ budget, Pareto recovers but not through diversity:}
NSGA-II-1000 achieves IGD 0.030 and Tier~1 on 91\% of datasets.
EZR-1000 matches this exactly (IGD 0.031, Tier~1 on 90\%),
confirming that the improvement is additional coverage reaching
the good region, not the diversity mechanism becoming useful,
consistent with the RQ2 finding.

\paragraph{The frontier is not degenerate:} All methods produce a median of 5.0 non-extreme balanced solutions. Hence, EZR delivers actionable
trade-off options similar to Pareto methods at 1000 evaluations, at
one-fifth the cost.

\begin{tcolorbox}[colback=gray!5, colframe=gray!40, boxrule=0.4pt,
  arc=1.5pt, left=4pt, right=4pt, top=2pt, bottom=2pt]
\small\textbf{RQ3.} At equal budget EZR beats Pareto methods
on their own metrics (IGD 0.040 vs.\ 0.085/0.066; Tier~1 IGD
65\% vs.\ 20\%/26\%), delivering the same 5.0 median trade-off
options at one-fifth the cost. Under this specific landscape, diversity mechanisms do not build
coverage under constrained budgets.
\end{tcolorbox}
\vspace{-0.4cm}
\section{Discussion}
\label{sec:discussion}
\vskip -0.1cm
\subsection{A Unified Structural Explanation}
\vskip -0.1cm
All three research questions share one explanation. SE Pareto fronts are rare and doubly concentrated being tight in decision space and clustered near the low-D2H region. Hence, diversity mechanisms waste evaluations on sparse extremes, and as a result, random sampling already beats Pareto methods in 68-71\% of datasets (RQ1). EZR reaches a stable operating level by evaluation 150-200, while Pareto methods need $3.3-3.4\times$ and SMAC needs $5 \times$ that budget just to match it (RQ2). Because EZR's trajectory naturally accumulates in this same region, it also beats Pareto methods on their own frontier metrics at equal budget (RQ3). The root cause is simple: methods that wander spend most of their budget outside where the good solutions are.

The random baseline's competitive performance (68–71\% W+T against Pareto methods) does not suggest MOOT tasks are intrinsically easy. Rather, it confirms the landscape structure: when the Pareto region is a tight 0.6\% island, any method, even an uninformed one, has a reasonable chance of landing near it with 200 evaluations. Diversity mechanisms actively reduce this chance by directing evaluations away from the concentrated good region toward sparse extremes. EZR's directed search (97\% W+T over Random) shows that the tasks are not trivial; they do reward informed exploitation, just not frontier-wide diversity.

\vspace{-0.3cm}
\subsection{Actionable Trade-off Navigation via Decision Trees}
\begin{wrapfigure}{r}{0.3\textwidth}
\vspace{-10pt}
\scriptsize
\begin{verbatim}
win
===
.67   baseline
.63   KLOC <= 227.23                 
.59   |   KLOC <= 93.52              
.54   |   |   RELY > 2.51
.62   |   |   RELY <= 2.51           
.60   |   |   |   PCAP <= 4.46       
.67   |   |   |   PCAP > 4.46        
.67   |   KLOC > 93.52               
.63   |   |   KLOC <= 212.88         
.58   |   |   |   ACAP <= 4.38       
.69   |   |   |   ACAP > 4.38        
.74   |   |   KLOC > 212.88          
.72   KLOC > 227.23                  
.68   |   RESL > 3.82               
.64   |   |   RELY > 2.42           
.71   |   |   RELY <= 2.42           
.76   |   RESL <= 3.82               
.70   |   |   RUSE <= 4.37          
.66   |   |   |   PLEx <= 3.19      
.76   |   |   |   PLEx > 3.19        
.82   |   |   RUSE > 4.37           
\end{verbatim}
\vspace{-5pt}
\caption{EZR tree on XOMO\_OSP. \texttt{win} = performance gain.
As shown at top, before optimization, the win=67. 
Best wins were achieved with reducing lines of code and improving
reliability.}\label{fig:ezr-tree-output}
\end{wrapfigure}
A Pareto frontier shows which outcome trade-offs are achievable, and the underlying archive records the configurations that produce them. However, the archive alone does not identify which parameters matter most or within what bounds to reliably stay in the good region. A decision tree
trained on Best versus Rest fills this gap at no additional
evaluation cost, it identifies the few master variables and bounds
that produce low-D2H solutions~\cite{Menzies2025DataLight,
chen2026promisetune}.

For example, Figure~\ref{fig:ezr-tree-output},  shows the very small tree learned 
from the XOMO\_OSP. Here, our methods selected parts of the $x$ space where, compared to the baseline,
defects reduced by $\frac{5917}{1258}\approx 470\%$ and the development effect by  $\frac{656}{199}\approx329\%$.

The data set used to generate  Figure~\ref{fig:ezr-tree-output}
has 26 $x$ attributes. Yet, this tree only used 7. This pattern
repeats across all our data sets:  i.e. as seen in
Figure~\ref{fig:treecomplexity}, our methods never build trees with more than 10 variables. Such small trees can be manually browsed and used to discover how little we need to change the input in order to nudge  a result into another leaf. We argue that this answers  Rudin's call~\cite{rudin2019stop} for  using interpretable
models in high-stakes decisions. 

Note that, in principle, our trees could
be trained on any Pareto front, but that requires expensive
frontier search first. EZR's Best set  delivers equivalent guidance (in the form of very small trees) much
faster.
\vskip -0.3cm
\subsection{Practical Recommendations}
\vskip -0.2cm
Start with EZR at 200 evaluations: it matches or beats Pareto methods 
in 84--85\% of tasks, and its search trajectory already outperforms 
NSGA-II and SPEA2 on IGD and HV at equal budget. The complementary 
decision tree identifies which parameters to adjust and within what 
bounds - guidance no Pareto frontier can provide. If more 
objective-space coverage is needed, escalate to EZR at 1000 
evaluations, which matches NSGA-II-1000 on frontier coverage 
(IGD 0.031 vs 0.030) at orders of magnitude lower runtime. Pareto 
methods remain a reasonable choice when practitioners have strongly 
asymmetric preferences or prior domain knowledge suggests a genuinely 
multimodal front.
\section{Threats to Validity}
\label{sec:threats}
\vspace{-0.2cm}

\textit{Internal validity.} Stochastic comparisons across many
datasets risk false positives. Twenty independent replicates and
Scott-Knott address this. The
random baseline provides a further check: if diversity maintenance
helped, random sampling would not match Pareto methods at equal
budget.

\textit{External validity.} All datasets are tabular and
pre-labeled, excluding online
optimization and corresponding combinatorial constraints similar to many other optimization studies on tabular data \cite{chenli2023weights, Amiraliminimaldata}. Consistency
across all five domains reduces domain-specific risk, though
generalization beyond these settings requires further study. EZR's pool-based design requires a pre-labeled or exhaustively enumerable candidate set, which excludes purely online settings (e.g., compiler flag tuning where each configuration must be physically built and run). In such settings, the unlabeled pool can be replaced by a random pre-sample; the centroid-based selection still applies, but initial diversity of that pre-sample becomes important. Extending EZR to true online optimization is future work.

\textit{Construct validity.} D2H assumes equal-weight aggregation
as the default when preferences are unknown~\cite{hwang1981methods,
zeleny1982multiple}. This is appropriate given that Pareto
solutions lie tightly closer to the ideal than non-Pareto
ones across 94\% datasets in our case. IGD and HV results further confirm this landscape structure.

\textit{Algorithm configuration.} The algorithm parameters may not represent
their best performance. For NSGA-II and SPEA2, population
sizes of 10 and 20 for 200 and 1000 evaluation budgets were used respectively,
following evidence that smaller populations with more generations improve
performance~\cite{hort2021offspring}, giving Pareto methods advantage. EZR's parameters are equally untuned, so Pareto methods are if
anything advantaged in this study. SMAC uses default surrogate,
found strongest among other global optimizers on MOOT~\cite{Menzies2025DataLight}, making its failure a conservative
result. NSGA-II and SPEA2 were selected as the most widely used Pareto
methods in SBSE~\cite{chenli2023weights}, reflecting standard
community practice.
\vspace{-0.4cm}
\section{Related Work}
\label{sec:related}
\vspace{-0.2cm}

\paragraph{Pareto-based SBSE:} Pareto-based search is a dominant SBSE paradigm \cite{harman2012search},
applied across release planning~\cite{bagnall2001next}, 
hyperparameter optimization~\cite{tantithamthavorn2016automated}, configuration tuning \cite{chenli2023weights}, and process
modeling~\cite{chen2018beyond}, with NSGA-II~\cite{deb2002fast} and
SPEA2~\cite{zitzler2001spea2} as the standard algorithms and
NSGA-III~\cite{deb2014nsgaiii} and MOEA/D~\cite{zhang2007moead}
extending this to many-objective and decomposition-based settings.
Li et al.~\cite{li20} provide standard evaluation guidance
using IGD and HV, which our RQ3 follows. Chen and
Li~\cite{chenli2023weights} showed Pareto search wins given sufficient
convergence budget. We complement this by analysing where and why
Pareto methods fail across 100+ SE datasets under sub-convergence
budgets, analyzing the objective-space structure that makes
diversity maintenance wasteful in expensive SE settings.

\paragraph{Lightweight optimizers:}
Lightweight and sampling-based optimizers have shown competitive 
performance in SE at small budgets. SWAY~\cite{chen2018sampling} showed 
hyperplane-based sampling competes with Pareto methods; 
FLASH~\cite{nair2018flash} found near-optimal configurations in far fewer 
evaluations than model-based methods; and LITE~\cite{Menzies2025DataLight} 
demonstrated contrastive active learning suffices for SE analytics. 
PromiseTune~\cite{chen2026promisetune} suggests why: few master 
variables govern SE solution quality. Menzies and 
Ganguly~\cite{Menzies2025DataLight} further show that SE problems 
collapse into surprisingly few occupied regions. These results were 
limited to single-objective, non-Pareto  or small-scale settings. We extend the 
question to 100+ multi-objective tasks, with a comparison against 
Pareto methods and global Bayesian search under realistic evaluation 
budgets, showing how, when and why these fail.

\paragraph{Global Bayesian optimization and explainability:}
Global Bayesian optimization methods, including SMAC~\cite{hutter2011smac},
TPE~\cite{bergstra11TPE}, and DEHB~\cite{awad2021dehb}, use surrogate
models to balance exploration and exploitation across the decision space.
SMAC is included as stronger than two other methods TPE an DEHB, recently benchmarked
on MOOT~\cite{Menzies2025DataLight}, making its failure here a
conservative result. The decision-tree explainability step builds on
Rayegan and Menzies~\cite{Amiraliminimaldata}, which show that shallow Best-vs.-Rest
trees produce actionable configuration guidance connecting directly to
the levers practitioners can turn~\cite{chen2026promisetune,
Menzies2025DataLight}.
\vspace{-0.4cm}
\section{Conclusion}
\vspace{-0.2cm}
Across over 100 SE optimization tasks, Pareto frontier exploration and global
Bayesian search fail to justify their cost under realistic evaluation
budgets. Diversity maintenance does not merely fail to help - it
actively displaces useful evaluations: random sampling beats Pareto
methods in 68--71\% of datasets, and EZR wins or ties in 84--85\%
while running orders of magnitude faster and outperforming both
NSGA-II and SPEA2 on their own frontier coverage metrics (IGD, HV).
The cause is structural: SE Pareto fronts live in a tiny concentrated region aligned with 
lower D2H, so a method
that zooms toward the best-vs-rest boundary finds them naturally
without frontier mapping or global exploration.

These results do not make Pareto methods obsolete. They remain the
right choice when practitioners have strongly asymmetric preferences
or when domain knowledge suggests a genuinely multimodal front -
the 15--16\% of datasets where EZR does not win or tie signals that
such cases exist. The most pressing open problem is a cheap
pre-screening test to identify these cases in advance, enabling
principled algorithm selection without paying the Pareto cost to
find out. Extending this evaluation to online optimization, hard
combinatorial problems, and many-objective settings where diversity
may be harder to avoid, are the natural next steps. Until then,
SE practitioners should zoom, not wander.
\vspace{-0.4cm}
\begingroup
\footnotesize
\bibliographystyle{unsrt}
\bibliography{ref}
\endgroup

\end{document}